\documentclass[5p,authoryear]{elsarticle}  % 2-column, nice
\usepackage{graphicx}
\usepackage{graphics}
\usepackage{natbib}
\usepackage{url}
\usepackage{epstopdf}
\usepackage{color}
\usepackage{bm}
\usepackage{times}
\usepackage{float}
\usepackage{lineno} % For submission
\newcommand{\icarus}{Icarus}

\def\lsim{\mathrel{\hbox{\rlap{\hbox{\lower4pt\hbox{$\sim$}}}\hbox{$<$}}}}
\def\gsim{\mathrel{\hbox{\rlap{\hbox{\lower4pt\hbox{$\sim$}}}\hbox{$>$}}}}

\def\ft2pi{\frac{1}{\left(\sqrt{2\pi}\right)^{3}}}

\def\d3{\mbox{d}}

\begin{document}

\title{Lunar polar craters -- icy, rough or just sloping?}

\author[icc]{Vincent R. Eke\corref{cor1}}
\ead{v.r.eke@durham.ac.uk}
\author[icc]{Sarah A. Bartram}
\author[icc]{David A. Lane}
\author[icc]{David Smith}
\author[ames]{Luis F. A. Teodoro}
\cortext[cor1]{Corresponding author}

\address[icc]{Institute for Computational Cosmology, Department of Physics, Durham University, Science Laboratories, South Road, Durham DH1 3LE, U.K.}
\address[ames]{BAER, Planetary Systems Branch, Space Science and Astrobiology Division, MS: 245-3, NASA Ames Research Center, Moffett Field, CA 94035-1000, U.S.A.}

\begin{abstract}

Circular Polarisation Ratio (CPR) mosaics from Mini-SAR on
Chandrayaan-1 and Mini-RF on LRO are used to study craters near to the
lunar north pole. The look direction of the detectors strongly affects
the appearance of the crater CPR maps.
Rectifying the mosaics to account for parallax also
significantly changes the CPR maps of the crater interiors. It is
shown that the CPRs of crater interiors in unrectified maps are biased to
larger values than crater exteriors, because of a combination of the
effects of parallax and incidence angle. Using the
LOLA Digital Elevation Map (DEM), the variation of CPR with angle of
incidence has been studied. For fresh craters, CPR $\sim 0.7$
with only a weak dependence on angle of incidence or position interior or just
exterior to the crater, consistent with dihedral scattering from
blocky surface roughness. For anomalous craters, the CPR interior to the crater
increases with both incidence angle and distance from the crater
centre. Central crater CPRs are similar to those in the crater
exteriors. CPR does not 
appear to correlate with temperature within craters. Furthermore, the
anomalous polar craters have diameter-to-depth ratios that are lower
than those of typical polar craters. These results strongly suggest
that the high CPR values in anomalous polar craters are not providing
evidence of significant volumes of water ice. Rather, anomalous
craters are of intermediate age, and maintain sufficiently steep sides
that sufficient regolith does not cover all rough surfaces.

\end{abstract}


\maketitle

\section{Introduction}

Knowing the quantity of water ice that is squirreled away in permanently
shaded lunar polar cold traps will constrain models of volatile
molecule delivery and retention. It is also of interest as a potential
resource for future explorers. The seminal work of 
\cite{wat61} introduced the possibility of water ice accumulations in
regions so cold, beneath $\sim 110$K, that ice would be stable against
sublimation for billions of years. Using the Lunar Prospector Neutron
Spectrometer (LPNS), \cite{feld98} showed that there were
concentrations of hydrogen at polar latitudes to the $70$ cm depths
probed by the neutrons. \cite{eke09} showed, with a pixon image
reconstruction algorithm that sharpened the LPNS hydrogen map, that
the excess polar hydrogen was preferentially concentrated into the
permanently shaded regions. However, while suggestive, the level of
$\sim 1$ wt$\%$ Water 
Equivalent Hydrogen (WEH), inferred from the models of \cite{law06},
was still not sufficiently high to prove that the 
hydrogen needed to be present as water ice. Only with the LCROSS
impactor \citep{cola10} did it become clear that water ice did indeed
exist, in a small region within Cabeus, at a level of a few per
cent by mass within the top metre or two of regolith. The hydrogen
maps produced from the LPNS by \cite{luis10} implied that there may
well be significant heterogeneity between permanently shaded polar
craters, so the LCROSS result should not be assumed to apply to all of
these cold traps.

Infra-red spectroscopy of the sunlit lunar surface has shown not only
absorption by surficial water and hydroxyl \citep{pie09,clark09}, but
also that these molecules are mobile across the surface depending upon
the time of lunar day \citep{sun09}. This supports the idea of a lunar
``water cycle'' of the sort envisaged by \cite{but97} and \cite{cv00},
but major uncertainties remain in our understanding of the efficiency
with which cold traps protect the volatiles that they receive \citep{cv03}.

The Lyman Alpha Mapping Project (LAMP) instrument on LRO has shown,
using radiation resulting from distant stars or scattering of the
Sun's Ly $\alpha$ from interplanetary hydrogen atoms,
that permanently shaded polar craters typically have a low far-UV albedo 
\citep{glad12}. These results are consistent with $1-2\%$ water frost
in the upper micron of the regolith of the permanently shaded
regions, with the observed 
heterogeneity between different craters perhaps implying a sensitivity
to local temperatures. Knowing how heterogeneous the water ice
abundance is would provide insight into which physical 
processes are most relevant for determining volatile retention.

Another widely-used remote sensing technique with the potential to
provide information about both the composition and structure of
near-surface material is radar \citep{cambook}. This often involves
sensing the polarisation 
state of the reflected radiation when circularly polarised radio waves
are transmitted
towards a surface. The dielectric properties of the materials present,
surface roughness, including rocks and boulders, composition and size of
any buried materials within the regolith and the depth of regolith
above bedrock could all affect the returned signal. For $13$ cm
radiation, the dielectric properties of regolith are such that the
upper few metres of the surface can be probed by radar
measurements. Given the complex 
nature of the scattering problem, it can be difficult to know what to
infer from radar data without additional insights into the likely
surface composition or structure. The most frequently used way of
characterising the returned signal is to take the ratio of powers in
the same sense (as transmitted) to the opposite sense of circular
polarisation, namely the circular polarisation ratio, or CPR. A CPR of
zero would be expected for specular reflection from a medium with
higher refractive index, whereas higher CPR values can result from
multiple scattering, which may imply the presence of a low-loss medium
such as water ice making up the regolith.

Radar observations of Europa, Ganymede and Callisto showed
surprisingly high CPR values of $\sim 1.5$ \citep{cam78,ost92}. The
low densities of these satellites were indicative of them having icy
compositions. The temptation to associate high CPR values with ice
increased when observations of the polar regions of Mercury showed
that high CPR regions were associated with permanently shaded craters,
within which temperatures could be low enough for water ice to be
stable against sublimation \citep{harm94}. Recent results from
MESSENGER's neutron spectrometer \citep{law13} support this conclusion.

It is less clear what should be inferred from radar observations of the 
Moon about the presence of water ice in permanently shaded craters. 
The Clementine mission transmitted circularly polarised radio waves
into the lunar polar regions, with the reflected flux measured on
Earth. An increase in same-sense polarised power at zero phase angle
was interpreted by \cite{noz96} as possible evidence for constructive
interference from waves taking reversed routes involving multiple
scattering within an icy regolith. This coherent backscatter
opposition effect \citep[CBOE][]{hapke} is one physical process that would
produce high CPR values. However, \cite{stac97}, \cite{st99} and \cite{cam05}
showed that high CPR could also result from surfaces that were rough
on scales within an order of magnitude in size of the $13$ cm radar
wavelength, which would help to explain why at least some of the high
CPR regions occurred in clearly sunlit locations where water ice would
not exist in significant amounts.

In parallel with the acquisition of remote sensing radar data, various
models have been constructed to help to interpret the CPR measurements.
Descriptions of the scattering mechanisms relevant to the problem are
given by \cite{cambook,cam12}. An empirical two-component model was
developed by \cite{thomp11} with a view to decoding CPR data from the
Mini-SAR and Mini-RF instruments on Chandrayaan-1 and LRO respectively.
The most physically motivated modelling to date was carried out by
\cite{fa11} who used vector radiative transfer theory to follow the
polarisation state of the input electromagnetic radiation. While their model
did not include multiple scattering, so had no CBOE, it did predict the
impact of incidence angle, regolith thickness, buried rocks and
surface roughness on the returned signal. They found that the
similarity in dielectric permittivity between ice and a silicate
regolith would make it difficult to identify ice mixed into
such a regolith.

The wealth of recent information returned from lunar missions provides
the possibility of discriminating between the different reasons for
high CPR regions on the lunar surface. \cite{spu10} used the
north pole CPR mosaic from the Mini-SAR instrument on Chandrayaan-1 to 
show how fresh craters showed high CPR both inside and out, whereas
a set of `anomalous' polar craters had high interior CPRs without any
corresponding enhancement just outside their rims. If meteorite
bombardment 
removed roughness at a similar rate inside and outside these craters
then this is suggestive that something other than roughness was
responsible for the anomalously high CPRs inside these
craters. That something could be water ice. Using Mini-RF data from
LRO, \cite{spu13} argued that
the abundance of anomalous craters was much greater near to the lunar
poles than at lower latitudes, with the implication that temperature
might be an important variable in determining the CPR in these craters.

More recently, \cite{fa13} studied examples of both polar and non-polar 
fresh and anomalous craters using data from the Mini-RF Synthetic
Aperture Radar instrument on
board LRO, finding polar and non-polar anomalous craters to have
indistinguishable distributions of pixel CPR. Given that water ice is
not the reason for the non-polar crater interiors having anomalously high 
pixel CPR values, why should it be necessary for the high pixel CPR
values in anomalous polar
craters? Furthermore, \cite{fa13} used LROC images to see boulders
within, and not outside, the non-polar anomalous crater. Despite the
mismatch in scales between the $>$1-2 m-sized rocks and the $13$ cm
radar wavelength, the model of \cite{fa13} shows that dihedral
scattering from such rocks can still significantly increase the CPR.
This provides a
potential reason for the anomalous crater CPR distributions and evidence
for some differential weathering from the crater interior to its
exterior. Unfortunately, the lack of illumination into the floors of
the polar craters precluded such a detailed investigation of rockiness
being carried out in these locations. In their detailed study of 
Shackleton crater, \cite{thom12} found that ``Mini-RF observations indicate 
a patchy, heterogeneous enhancement in CPR on the crater walls whose strength
decreases with depth toward the crater floor.'' While placing an upper limit 
of $\sim 5-10$ wt\% H$_2$O ice in the uppermost metre of regolith, they conclude 
that the result ``... is most consistent with a roughness effect due to
less mature regolith present on the crater wall slopes.''

In this paper, the polar craters studied by \cite{spu10} will be
investigated using a combination of topography, radar and temperature
data sets, with a view to determining what is responsible for the
anomalous polar craters, and is anything special about their cold floors.
Section~\ref{sec:data} contains descriptions of the various data sets
that will be employed and the set of polar craters to be
studied. Results concerning the variation of CPR with incidence angle
and position within the crater, as well as a simple model showing the
impact of parallax in the range measurement, are contained in
Section~\ref{sec:res}. What these CPR
measurements imply about the presence of polar water ice are discussed
in Section~\ref{sec:disc}, and conclusions drawn in
Section~\ref{sec:conc}. 

\section{Data}\label{sec:data}

A number of different lunar data sets, available from the Geosciences Node of
NASA's Planetary Data System
(PDS\footnote{http://pds-geosciences.wustl.edu}), will be used. This
section describes them briefly, as well as providing details of the
set of north polar craters to be studied.

\subsection{LOLA Topographical data}\label{ssec:lola}

The polar stereographic Lunar Orbiter Laser Altimeter (LOLA) Digital
Elevation Map (DEM) for the north pole, with 
a pixel size of $80$ m, is used in this study \citep{smith10}. These
data are used for finding craters using the algorithm defined in the
Appendix, which returns crater locations, diameters ($D$) and depths
($d$), and also to determine surface normals and hence radar angles of
incidence for the Synthetic Aperture Radar (SAR) observations. 

\subsection{Synthetic Aperture Radar data}\label{ssec:radar}

Both the S-band ($12.6$ cm wavelength) CPR and reflected power
(characterised through the first element of the Stokes vector, $S_1$)
polar stereographic mosaics for the 
Mini-SAR instrument on Chandrayaan-1 \citep{spu09} and Mini-RF on LRO
\citep{noz10} are used here. These instruments use a hybrid polarity
architecture \citep{raney07}, emitting circularly polarised radio waves and
receiving two orthogonal linear polarisations coherently, enabling the
Stokes vector of the returned signal to be fully reconstructed. The PDS
mosaics of CPR and $S_1$ provide measurements with a pixel size of $75$
m for Mini-SAR and $\sim 118$ m for Mini-RF down to a latitude of
$\sim 70^\circ$. Both of these instruments 
were side-facing, relative to the direction of spacecraft motion, with
Mini-SAR having a nadir angle of $\sim 33^\circ$ and Mini-RF 
$\sim 48^\circ$. The currently available
mosaics are neither controlled, to take into account the imperfect
knowledge of the spacecraft trajectory, nor orthorectified to tie the
images to an underlying base map such as that provided by the LOLA
DEM. Orthorectification involves removing distortions in the inferred
range distance, perpendicular to the direction of spacecraft motion,
resulting from height variations in the topography affecting 
the return times of the radar pulses \citep{kirk13,cambook}. The
impact of this radar parallax effect is significant and will be
considered in detail in this paper. 
These factors mean that the Mini-SAR and Mini-RF mosaics can be spatially
offset from the base map set by the LOLA DEM by up to 
$\sim 5$ km and $\sim 2$ km respectively. The Mini-RF mosaic is a
mixture of left- and right-looking measurements, with most pixels being
assigned the latest right-looking observation, with $\sim 5\%$ of pixels
being left-looking (R. Kirk, private communication). Consequently, the
Mini-RF mosaic will not be used for the quantitative analysis towards
the end of this paper. It should be
noted that near to the poles, right-looking does not imply
east-looking. For instance, when the detector is at the north pole,
right-looking corresponds to facing south.

\subsection{Diviner data}\label{ssec:div}

\begin{figure*}
\begin{center}
\includegraphics[trim=2cm 1cm 1cm 4cm,clip=true,width=2.3\columnwidth]{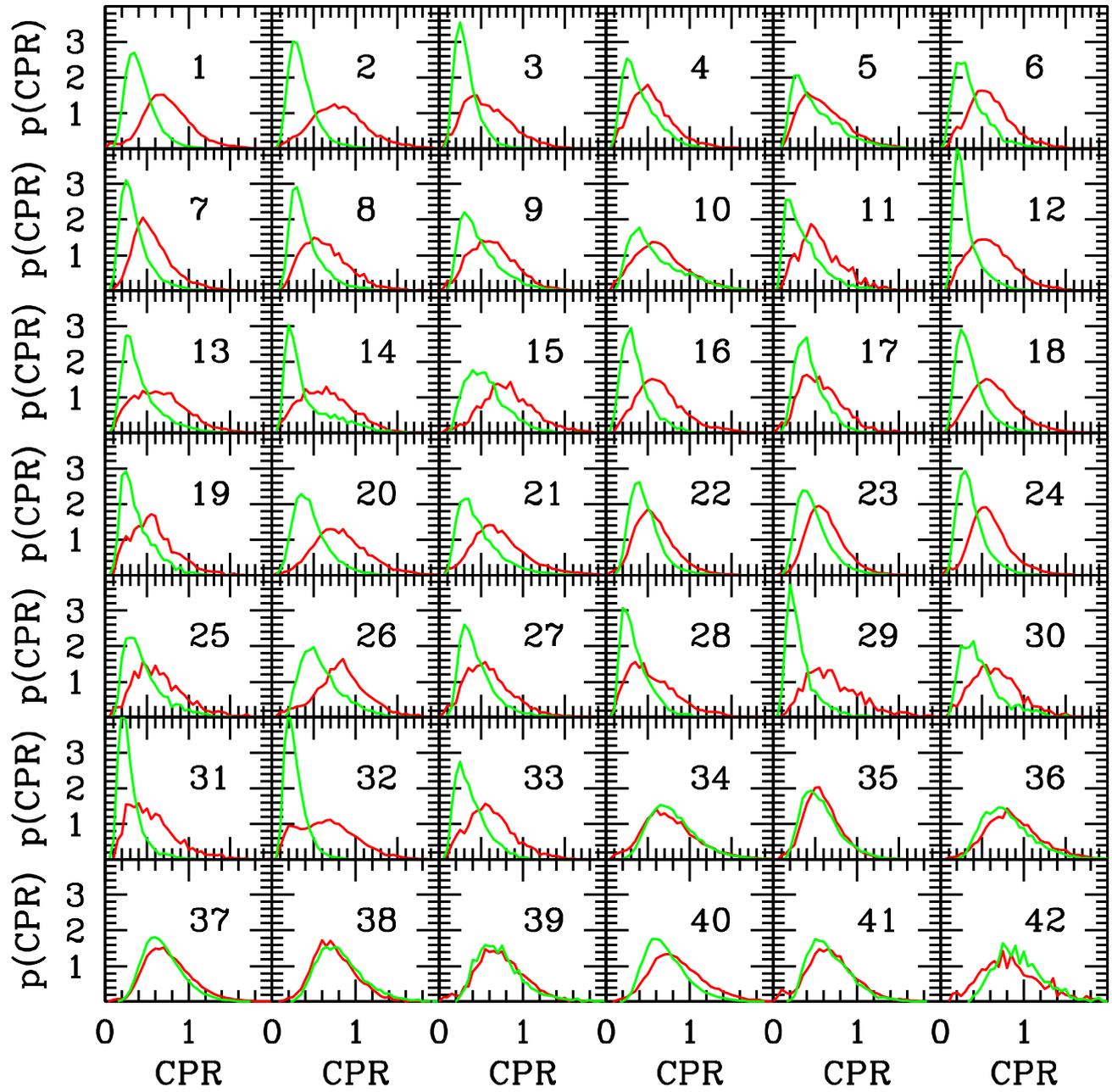} % 2-column
\end{center}
\vspace{-5cm}
\caption{The distributions of pixel CPR for the 42 craters considered,
measured from the unrectified Mini-SAR mosaic. Pixels interior to the
crater are shown in red and those with radii satisfying 
$1<r/r_{\rm c}<1.5$ are shown in green. The anomalous craters (numbers
1-33) have significantly different interior and exterior pixel CPR
distributions, with the interior distribution skewed to higher values
than is seen from regions just outside the crater rim. The fresh
craters (numbers 34-42) have very similar 
interior and exterior CPR distributions.}
\label{fig:allcdist}
\end{figure*}

The Diviner infra-red radiometer on board LRO has measured
fluxes from the lunar surface in nine different spectral bands,
allowing surface temperatures to be inferred. From these data, with
a model to account for the variation in solar illumination over time,
maps of average and maximum temperatures can be calculated
\citep{paige10b}. Given the exponential dependence of both water
molecule diffusion and sublimation rates on
temperature, the map of maximum temperature is likely to be most
relevant to the distribution of polar water ice and is used
here. These $T_{\rm max}$ values are provided in a set of triangular
pixels poleward of $75^\circ$ latitude, with a spatial resolution of
$\sim 500$ m.

\subsection{The crater set}\label{ssec:crat}

\begin{table}\label{tab:loc}
\begin{center}
\caption{Radii and locations for craters used in this
study. Longitudes and latitudes are given in degrees. Different
locations are used for the two radar data sets on account of the
available mosaics not having been tied to the LOLA base
map. Uncertainties on the locations are $\sim80$ m, $\sim150$~m and
$\sim250$ m for LOLA, Mini-SAR and Mini-RF respectively.}
\footnotesize{
\begin{tabular}{l|c|c|c|c}\hline
Crater \# & Radius  & LOLA & Mini-SAR & Mini-RF \\
& $r_{\rm c}$/km &  (lat,\,lon) &  (lat,\,lon) & (lat,\,lon) \\  \hline
1 & 6.0 & 79.04, -148.4 & 78.89, -149.0 & 78.98, -148.4 \\
2 & 4.3 & 84.05, -156.4 & 83.88, -157.4 & 84.02, -156.5 \\
3 & 3.2 & 80.17, -124.6 & 80.07, -124.7 & 80.13, -124.7 \\
4 & 3.8 & 80.45, -122.6 & 80.33, -122.9 & 80.41, -122.8 \\
5 & 3.6 & 85.78, 25.2 & 85.68, 25.4 & 85.73, 24.9 \\
6 & 2.9 & 85.75, 43.6 & 85.69, 44.7 & 85.72, 43.5 \\
7 & 5.3 & 86.99, 28.6 & 87.08, 30.1 & 86.94, 28.2 \\
8 & 2.7 & 88.08, 39.9 & 88.10, 43.9 & 88.05, 40.6 \\
9 & 3.4 & 87.73, 16.9 & 87.66, 19.0 & 87.74, 15.7 \\
10 & 2.9 & 87.97, 29.9 & 88.21, 29.4 & 87.97, 28.2 \\
11 & 1.7 & 89.13, 59.5 & 89.09, 69.8 & 89.10, 60.9 \\
12 & 3.3 & 88.19, 63.4 & 88.20, 67.4 & 88.15, 63.5 \\
13 & 2.8 & 86.59, 93.2 & 86.47, 93.6 & 86.56, 92.6 \\
14 & 2.5 & 88.75, 47.1 & 88.69, 52.3 & 88.72, 48.0 \\
15 & 1.9 & 81.80, -110.0 & 81.65, -111.1 & 81.75, -110.0 \\
16 & 2.4 & 82.67, -83.6 & 82.53, -84.6 & 82.62, -83.7 \\
17 & 2.0 & 82.75, -80.8 & 82.62, -81.9 & 82.70, -80.9 \\
18 & 8.7 & 80.26, -50.1 & 80.19, -50.3 & 80.22, -50.2 \\
19 & 1.9 & 86.31, -89.1 & 86.17, -90.1 & 86.27, -89.4 \\
20 & 4.1 & 87.14, -86.3 & 86.99, -87.4 & 87.17, -86.1 \\
21 & 4.8 & 81.65, -23.9 & 81.58, -24.1 & 81.59, -23.9 \\
22 & 3.8 & 85.14, -166.7 & 84.97, -167.9 & 85.11, -166.8 \\
23 & 9.6 & 87.98, -52.2 & 87.91, -52.7 & 88.00, -51.7 \\
24 & 5.3 & 83.75, -13.8 & 83.67, -14.4 & 83.71, -14.0 \\
25 & 2.0 & 86.19, -177.5 & 86.01, -178.8 & 86.14, -177.6 \\
26 & 2.8 & 86.81, -13.9 & 86.72, -14.4 & 86.77, -14.6 \\
27 & 2.5 & 84.99, -2.0 & 84.90, -2.7 & 84.95, -2.2 \\
28 & 2.4 & 87.83, 113.0 & 87.67, 111.1 & 87.81, 112.3 \\
29 & 1.8 & 86.81, 116.1 & 86.80, 118.5 & 86.78, 115.4 \\
30 & 1.8 & 85.93, 111.7 & 85.80, 111.4 & 85.90, 111.3 \\
31 & 1.5 & 85.43, 105.3 & 85.32, 105.3 & 85.40, 105.0 \\
32 & 5.4 & 81.15, 137.7 & 81.22, 138.3 & 81.12, 137.6 \\
33 & 2.3 & 82.12, 92.3 & 81.99, 91.7 & 82.09, 92.1 \\
34 & 6.5 & 81.45, 22.6 & 81.35, 22.6 & 81.40, 22.9 \\
35 & 4.7 & 84.86, 35.6 & 84.76, 35.7 & 84.81, 35.5 \\
36 & 2.3 & 87.69, 30.8 & 87.74, 33.9 & 87.68, 29.6 \\
37 & 9.8 & 82.42, -68.7 & 82.32, -68.7 & 82.38, -68.8 \\
38 & 2.7 & 84.48, -132.4 & 84.34, -133.1 & 84.44, -132.3 \\
39 & 1.6 & 81.62, -161.7 & 81.51, -161.4 & 81.58, -161.7 \\
40 & 6.4 & 84.82, -172.2 & 84.67, -173.0 & 84.79, -172.4 \\
41 & 2.8 & 80.93, 117.1 & 80.82, 117.4 & 80.88, 117.0 \\
42 & 1.2 & 86.16, 71.0 & 86.06, 71.6 & 86.12, 70.7 \\
\hline
\end{tabular}}
\end{center}
\end{table}

A set of polar craters was found by applying the algorithm described
in the Appendix to the LOLA $80$ m north pole stereographic DEM.
Briefly, this method involves finding depressions in the surface
by tracking to where `water', placed uniformly across the surface,
runs. Isolated `puddles' provide possible candidates for simple,
isolated craters that do not have significant sub-cratering. A
crater-shaped filter is run over the DEM in the vicinity of
sufficiently isolated depressions. This filter picks out circularly
symmetric concave regions with a circular convex rim. The best match
of the crater-shaped filter with the DEM defines the crater centre and
radius, $r_{\rm c}$, and the value of the filtered DEM provides a quantitative
measure of how crater-like each candidate is.

$42$ of the craters studied by \cite{spu10} were matched to
crater candidates in the LOLA DEM. Locations and radii are provided in
Table~1 for this set. Note that, because the Mini-SAR and
Mini-RF mosaics have not been orthorectified to the LOLA base map,
there are different crater centres for each of these data
sets. To determine the crater centres, their radii and
approximate locations are taken from the crater-finding algorithm. The
radar data are then visually aligned, matching the pattern of nearby
craters in the LOLA DEM to those visible in the CPR and $S_1$ maps.
In the radar data, anomalous and fresh craters show up as regions of
high CPR, with arcs of high $S_1$ on the far crater walls. The
accuracy with which this alignment can be used to estimate the
positions of the crater rims is approximately $2$ pixels, which is
$150$ m for the Mini-SAR data. This is less than $10\%$ of the crater
radius for almost all of the craters considered here. Having aligned
the rims of the craters in this way, the pre-rectification centre
locations are assumed to have the same uncertainty in position.
A few of the
craters studied by \cite{spu10} are not included in the sample of $42$
craters, either because they could not be confidently found in the CPR
maps, or  because their CPR and $S_1$ distributions did not allow a
clear centre to be inferred. 

Figure~\ref{fig:allcdist} shows probability distributions for pixel
CPR values measured from the Mini-SAR mosaic for the interiors and
exteriors of all $42$ craters. Craters $1-33$ represent the
``anomalous'' ones with exterior CPR values being typically lower than
interior ones, whereas numbers $34-42$ are fresh craters. For
reference, crater $2$ is the anomalous crater shown in figure 3 of
\cite{spu10}. 

\section{Results}\label{sec:res}

The different CPR distributions for pixels interior and exterior to
the polar anomalous craters are clearly seen in
Figure~\ref{fig:allcdist}. This section contains the results from a
more detailed analysis of what gives rise to these differences.

\subsection{Stacking craters}\label{ssec:stack}

\begin{figure}
\begin{center}
\includegraphics[trim=1cm 1cm 1cm 4cm,clip=true,width=0.95\columnwidth]{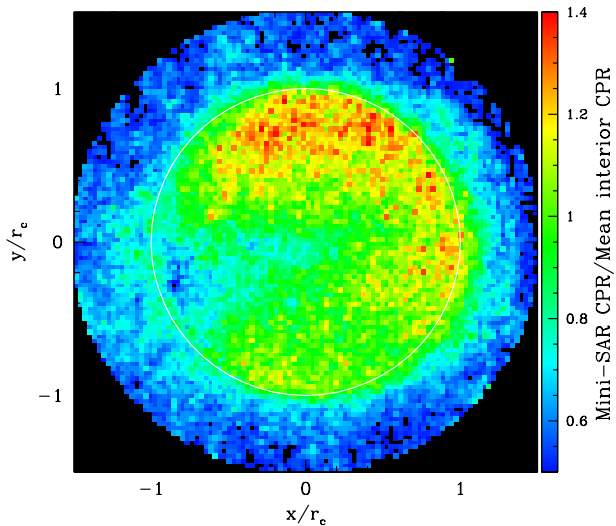}
\end{center}
\vspace{-2cm}
\caption{The stacked relative CPR map for the 33 anomalous
craters. Each crater map is divided by the mean pixel CPR interior
to the crater and rotated to have north at the top before they are
stacked together. The white circle represents the edge of the craters
contributing to the average.}
\label{fig:ch1stack}
\end{figure}

If the anomalously high interior CPR measurements in polar craters
were the result of significant deposits of water ice, then one might
expect to see a variation of CPR with the position within the
crater, reflecting varying insolation, temperature and hence water
ice stability \citep{vas99}. To enhance the signal-to-noise, all $33$
anomalous craters have been 
stacked together to produce the Mini-SAR CPR map shown in
Figure~\ref{fig:ch1stack}. The stacking process involves dividing each
pixel's CPR by the mean crater interior CPR and the distance from the
centre is expressed as a fraction of the distance to the crater's edge.
The map for each crater is rotated to have the north pole at the top,
and the final stacked map is the mean of these processed crater maps. 
It is apparent from the figure that the highest CPR is typically on
the poleward side of the crater, with a distinctive horseshoe pattern
of higher CPR around the crater walls. 

\begin{figure}
\begin{center}
\includegraphics[trim=1cm 1cm 1cm 4cm,clip=true,width=0.95\columnwidth]{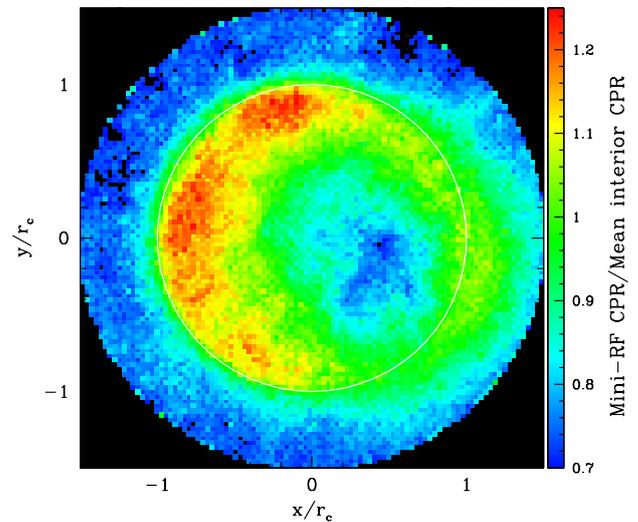}
\end{center}
\vspace{-2cm}
\caption{The stack of the 33 anomalous crater relative CPR maps using the LRO
Mini-RF unrectified mosaic.}
\label{fig:lrostack}
\end{figure}

Stacking the same $33$ anomalous craters together using the Mini-RF
mosaic gives rise to the CPR map in Figure~\ref{fig:lrostack}. Once
again a horseshoe-shaped high CPR region is seen, only in a different
part of the stacked crater. Given that the lunar surface will not have
changed significantly during the period between Mini-SAR and Mini-RF
data collection, it can be inferred that this difference reflects a
change in the viewing geometry, as anticipated by the model of
\cite{fa11} (see their figure 13).

\begin{figure}
\begin{center}
\includegraphics[trim=1cm 1cm 1cm 4cm,clip=true,width=0.95\columnwidth]{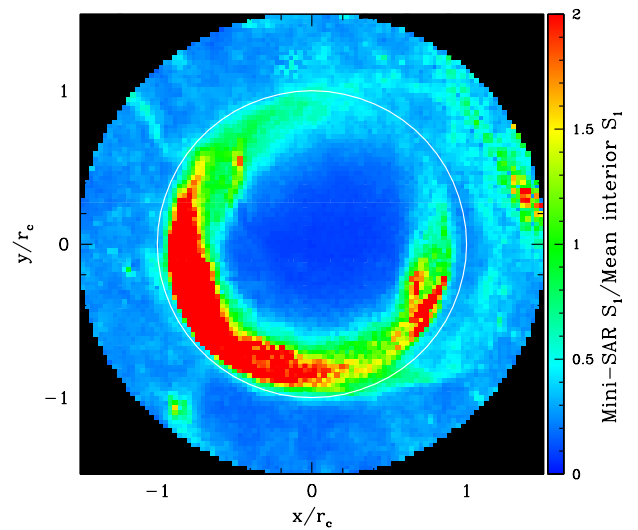}
\end{center}
\vspace{-2cm}
\caption{The stacked relative returned power, represented by the first
element of the Stokes vector, $S_1$, for observations of the 33
anomalous craters made by Mini-SAR. All craters are aligned so that
north points to the top of the image before stacking.}
\label{fig:s1cstack}
\end{figure}

\begin{figure}
\begin{center}
\includegraphics[trim=1cm 1cm 1cm 4cm,clip=true,width=0.95\columnwidth]{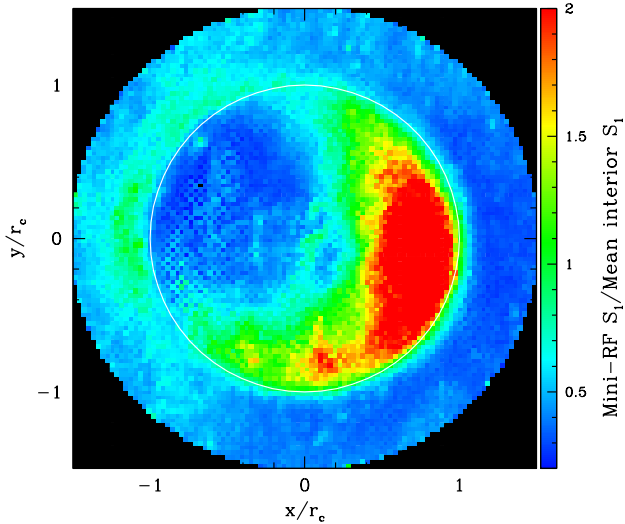}
\end{center}
\vspace{-2cm}
\caption{The equivalent of Fig.~\ref{fig:s1cstack} for the
  LRO Mini-RF $S_1$ mosaic.}
\label{fig:s1lstack}
\end{figure}

This conclusion is strengthened by
the corresponding stacked maps of the returned power shown in
Figures~\ref{fig:s1cstack} and~\ref{fig:s1lstack}, which are
determined from the $S_1$ mosaics. Higher returned power suggests the
transmitted radiation is nearer to normal incidence on the
surface. Consequently, there will be greater specular reflection and a
lower returned CPR. Thus, the highly reflective parts of the stacked
returned power maps correspond to the low parts in the CPR maps.
When the surface is viewed at larger angles
of incidence, the multiply scattered radiation becomes increasingly
important and the returned CPR increases while the returned power decreases.
The stacked crater maps shown in these figures all have north to the
top, but the radar look direction does not always have the same bearing
because the side-facing detector will change its look direction near
to the pole. In addition to having different look directions for the
different craters contributing to the stacked map, the incidence angle
in any given pixel will vary between craters as they have a variety of
diameter-to-depth ratios. Consequently, these stacked maps are for
illustrative purposes only, and all subsequent radar results treat the
craters individually, using a look direction 
inferred by determining the position of the maximum reflected power in that
crater's $S_1$ map.

From these figures, it is clear that the largest factor affecting the
CPR maps of these polar craters
is the angle of incidence of the observations. As the
Mini-RF mosaic includes both left and
right-looking measurements it will not be possible to infer an
appropriate, reliable single crater look direction from the mosaic, so
attention will now be focussed onto the Mini-SAR data.

\subsection{Slopes and parallax}\label{ssec:slopes}

Given that the angle of incidence is a complicating, and for the
purposes of learning about the lunar surface uninteresting, factor
driving the CPR distribution within the polar craters, it would be
good to remove its effect. While there have been models of how CPR
varies with angle of incidence \citep{thomp11,fa11}, a more robust
approach involves determining the dependence using the data
themselves.

Each crater has an $S_1$ map with a high spot that should be nearest to
normal incidence for the incoming radar. This is defined within a cone
of opening angle $20^\circ$ from the centre of the crater, and is used to
define the azimuthal look direction of the detector appropriate to
this particular crater. In combination
with the nadir angle of the detector, this provides a vector for the
incoming radiation. Finite differencing methods applied to the LOLA
DEM provide a local surface normal. The scalar product of these unit
vectors yields the cosine of the angle of incidence for each pixel in
each of the craters being considered. In this way, each
pixel CPR can be mapped to a corresponding angle of incidence.

\begin{figure}
\vspace{-1.5cm}
\begin{center}
\includegraphics[trim=0.5cm 1cm 1cm 0.5cm,clip=true,width=0.95\columnwidth]{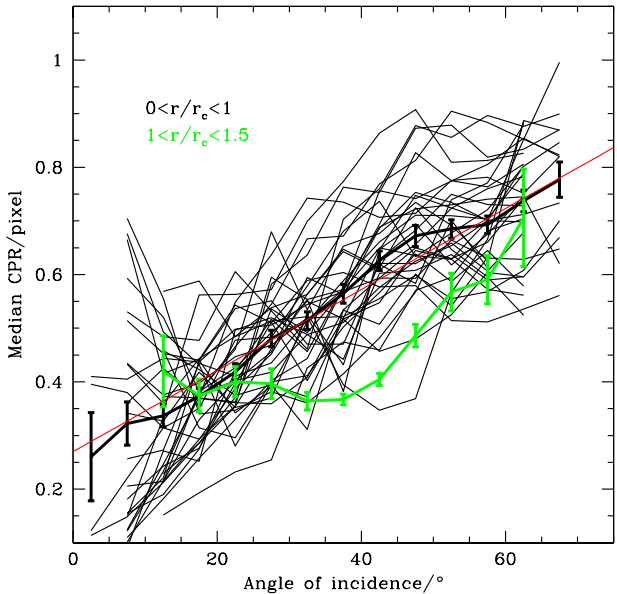}
\end{center}
\vspace{-2cm}
\caption{The variation of median CPR as a function of angle of
  incidence between the incident radar and the surface normal for the
  33 anomalous craters. The light black lines show the individual
  crater median pixel CPR curves, and the heavy black line is the
  median of these values.
  Error bars show an estimate of the statistical uncertainty on
  the inferred median based on the 16th and 84th percentiles of the
  distribution of CPR values from the individual craters at each angle
  of incidence and the assumption that this distribution is
  Gaussian. The heavy green line is the median over all craters for the
  crater exterior out to $1.5 r_{\rm c}$. Positions have been rectified
  to account for the parallax 
  prior to determining into which radial range they fall. The red line
  shows a straight line fit to the median interior CPR relation.}
\label{fig:angle}
\end{figure}

One final, but crucial, complication is to determine to which bit of the
surface does an unrectified Mini-SAR mosaic pixel correspond. The
effect of parallax in radar range measurements distorts the inferred
pixel position because the mapping of return signal time to distance
should account for variations in the height of the surface being
mapped \citep{cambook}. As the Mini-SAR crater positions have
been individually chosen such that the crater rims appear to line up
correctly (something that the stacked CPR and $S_1$ mosaics imply has
been done reasonably well), the mean altitude of the crater rim is set
as the reference height. All other points within $1.5r_{\rm c}$ of the
crater centre are then shifted a distance $p$ away from the detector
in the range direction using
\begin{equation}
\Delta h=p \tan\alpha,
\end{equation}
where $\Delta h$ represents the change in height, at the shifted
position, relative to the 
reference height, $p$ is the parallax, and $\alpha$ is the angle of
incidence of the radar \citep[see section 4.11 in][]{cambook}. 
An iterative procedure is necessary because the parallax displacements
depend upon the topography at to-be-determined positions in the DEM.
This shift moves
unrectified pixels within the crater having $\Delta h<0$ to
positions that are nearer to the detector (i.e. $p<0$). As a
consequence, equally spaced pixels in the distorted, unrectified map
preferentially sample the near crater wall at higher angles of
incidence.

Having determined which part of the LOLA DEM should be matched to each
pixel in the vicinities of the craters being considered, the
dependence of pixel CPR on the angle of incidence can be
determined. Figure~\ref{fig:angle} shows the median dependence of the
pixel values for each of the $33$ anomalous north pole craters being
considered here. The median of these curves is shown with the bold
black line, which can be well described by the linear fit
\begin{equation}
CPR(\theta)=0.27+0.68(\theta/90^\circ)
\label{cprtheta}
\end{equation}
where $\theta$ represents the angle of incidence in degrees. The
crater interior shows a strong trend of increasing CPR with
increasing angle of incidence, although the individual crater values
have a non-negligible scatter about this median relation. A bold green
line traces the median dependence for the $33$ crater exterior regions out
to $1.5r_{\rm c}$, and clearly shows lower CPR values for intermediate angles
of incidence than are typical inside these craters. 
While the exterior CPR does become more similar to the interior crater
values at high and low angles of incidence, it is possible that
this is a consequence of inaccuracies in defining the crater edges in
the Mini-SAR mosaic.

This measurement of the variation of CPR with angle of incidence could
contain dependencies on hidden surface properties that have not been
considered, but it serves as a useful starting point for constructing
a simple model with which to investigate just how important the
rectification process is. A model crater was created with diameter
$2r_{\rm c}=6$ km, and a diameter-to-depth ratio of $5.5$, typical of the
anomalous polar craters considered here.
The radial height profile, $a(x)$, with $x=r/r_{\rm c}$ being the radius in
terms of the crater radius, was defined via $y(x)=a(x)/r_{\rm c}$, where
\begin{equation}
%\footnotesize{
\hspace{-0.8cm}y(x)=\left \{
\begin{array}{ll}
y_0+\eta x^2 &{\rm if}~ x\leq x_1,\\
y_1+y_1'(x-x_1) &{\rm if}~ x_1\leq x\leq x_2,\\
y_2+\beta[(x_2-1)^2-(x-1)^2] &{\rm if}~ x_2\leq x\leq x_3,\\
y_3+\gamma[(x-x_4)^2-(x_3-x_4)^2] &{\rm if}~ x_3\leq x\leq x_4,\\
y_4 &{\rm if}~ x_4\leq x.
\end{array}
\right.
%s}
\end{equation}
$y_0$ represents the central depth divided by the crater radius, which
is just twice the reciprocal of the diameter-to-depth ratio, while
$y_n$ for $n>0$ is the value of $y$ evaluated at $x_n$. $y_1'$ denotes
d$y/$d$x$ evaluated at $x_1$. With the outer
boundary condition set as $y_4=-0.04$ at $x_4=1.5$ and the two inner
curvatures chosen to be $\eta=1$ and $\beta=2$, the requirements that the
function is continuous and differentiable sets the remaining constants
via
\begin{eqnarray}
x_1&=&\frac{1-\sqrt{1-\frac{y_0}{\eta}(1+\eta/\beta)}}{1+\eta/\beta}\\
x_2&=&1-\frac{\eta}{\beta} x_1\\
x_3&=&1-\frac{y_4}{\beta (x_4-1)}\\
\gamma&=&\frac{\beta (x_3-1)}{x_4-x_3}.
\end{eqnarray}
This cross-section for the model crater is shown in
Figure~\ref{fig:app2} and has a maximum smooth slope for the crater wall of 
tan$^{-1}y_1'\approx 23^\circ$.
A regular $75$ m grid of pixels was created out to $x_4=1.5$ from the
crater centre. Assuming that these pixels were unrectified, the
corresponding rectified positions in the crater were calculated, the
angles of incidence to the nominal detector with a nadir angle of
$33^\circ$ were inferred and CPR values were assigned according to 
equation~(\ref{cprtheta}).

\begin{figure}
\begin{center}
\includegraphics[trim=0cm 1cm 1cm 4cm,clip=true,width=0.95\columnwidth]{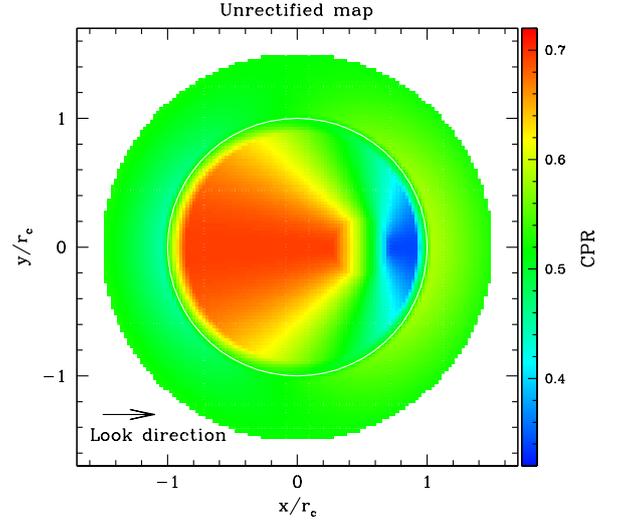}
\end{center}
\vspace{-2cm}
\caption{An unrectified CPR mosaic of a model crater with 
$r_{\rm c}=3$km, a diameter-to-depth ratio of $5.5$ and a rim height of
$0.04r_{\rm c}$. The model SAR is looking from the left with a look angle of
  $33^\circ$ and the mosaic has $75$m square pixels. }
\label{fig:modelu}
\end{figure}

\begin{figure}
\begin{center}
\includegraphics[trim=0cm 1cm 1cm 4cm,clip=true,width=0.95\columnwidth]{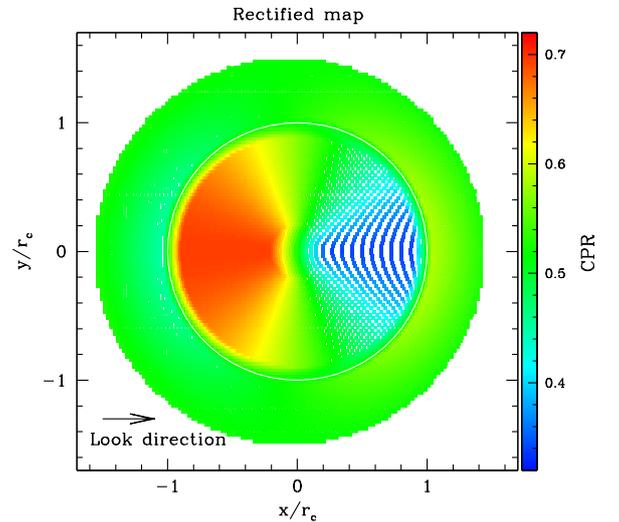}
\end{center}
\vspace{-2cm}
\caption{The rectified version of Fig.~\ref{fig:modelu}, with each
  coloured point showing the true position within the crater that it
  samples. White regions show parts of the crater into which none 
  of the unrectified mosaic pixels are mapped when the parallax correction
  moves pixels beneath the crater rim toward the detector. The colour
  relates directly to the angle of incidence at which the surface is
  viewed through equation~(\ref{cprtheta}).}
\label{fig:modelr}
\end{figure}

The resulting unrectified CPR mosaic is shown
in Figure~\ref{fig:modelu} from which it can be seen that the high CPR
values associated with the near wall, viewed at large angles of
incidence, occupy a significantly larger fraction of the crater
interior pixels than the more nearly normal incidence parts of the far wall.
Figure~\ref{fig:modelr} shows the same pixels shifted to the parts of
the crater that they actually sample. With the effect of parallax
removed from the map, it becomes apparent just how the pixels are
biased to measure the CPR of the near wall of the crater. Even with $75$ m
unrectified resolution of a $6$ km diameter crater, there are significant
parts of the far wall that are completely unsampled.

\begin{figure}
\begin{center}
\includegraphics[trim=0cm 1cm 1cm 3cm,clip=true,width=0.95\columnwidth]{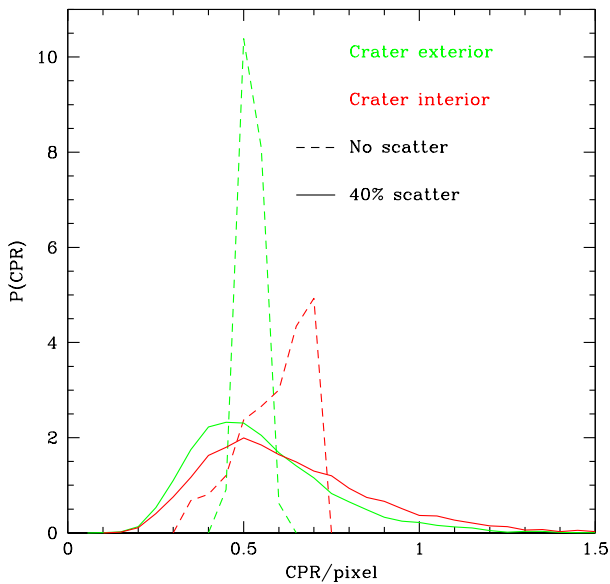}
\end{center}
\vspace{-2cm}
\caption{The distribution of pixel CPR values for the interior (red)
  and exterior (green) of the model crater. Dashed lines show results
  when no scatter is added in the model CPR value at a given angle of
  incidence, whereas the solid lines show the effect of including a
  $40$\% $1\sigma$ Gaussian scatter around the median value.}
\label{fig:modelcdist}
\end{figure}

The impact of this uneven sampling of the crater
on the probability distribution of pixel CPR values
is shown in Figure~\ref{fig:modelcdist}. Dashed red and green
lines show how the interior and exterior pixel CPR distributions can
look significantly different, despite both being drawn from an
identical relation for CPR as a function of angle of incidence. The
peak of the distribution shifts from a CPR of $\sim 0.5$ to $\sim
0.7$, as a result only of the bias caused by using a mosaic
uncorrected for the effect of parallax and the dependence of CPR on
angle of incidence. 
These pixel CPR distributions are much more sharply peaked than those in
Figure~\ref{fig:allcdist} that were measured for real craters using
the Mini-SAR mosaic. One way in which the distribution would be
broadened would be if there were significant statistical uncertainties
on the measurements. The solid lines in Figure~\ref{fig:modelcdist}
show that including a $40\%$ scatter in the assumed CPR at any
particular angle of incidence produces distributions that
look not unlike those from a few of the anomalous craters.

\begin{figure}
\begin{center}
\includegraphics[trim=0cm 1cm 1cm 3cm,clip=true,width=0.95\columnwidth]{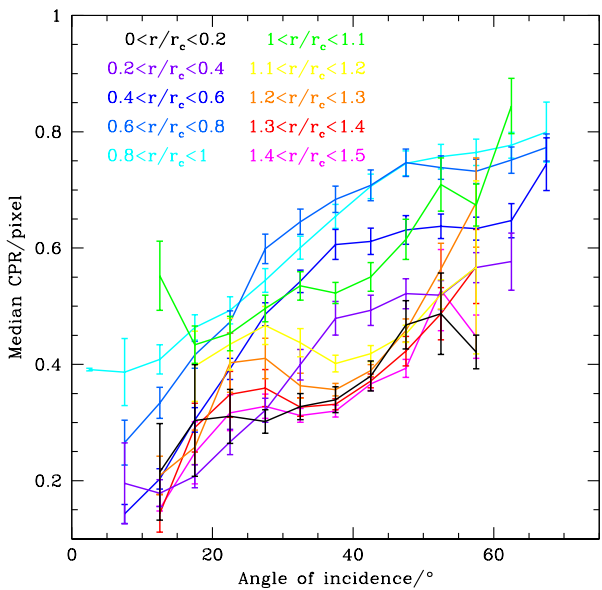}
\end{center}
\vspace{-2cm}
\caption{The variation of median CPR as a function of angle of
  incidence between the incident radar and the surface normal for the
  33 anomalous craters. Values show the median of the individual
  crater values that contribute to each increment of incidence
  angle. Error bars show an estimate of the statistical uncertainty on
  the inferred median based on the 16th and 84th percentiles of the
  distribution of CPR values from the individual craters at each angle
  of incidence and the assumption that this distribution is
  Gaussian. The different colours represent different radial ranges of
  pixels. Positions have been rectified to account for the parallax
  prior to determining into which radial range they fall.}
\label{fig:anglea}
\end{figure}

\begin{figure}
\begin{center}
\includegraphics[trim=0cm 1cm 1cm 3cm,clip=true,width=0.95\columnwidth]{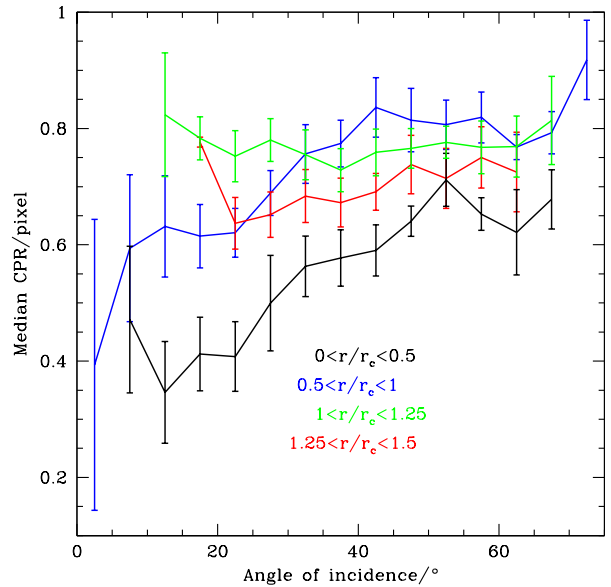}
\end{center}
\vspace{-2cm}
\caption{The equivalent of Fig.~\ref{fig:anglea} for the 9 fresh
  craters. Wider radial ranges are used to suppress statistical noise
  in the median CPR estimates.}
\label{fig:anglef}
\end{figure}

Is it reasonable that such large observational uncertainties exist?
This can be indirectly addressed by considering the variation in CPR
between adjacent pixels in the Mini-SAR mosaic. The root mean square
fractional difference in CPR varies only slightly across the whole
polar region, and typically has a value of $25-30\%$ in the vicinity
of the craters studied here. This represents an upper limit on the
size of the statistical uncertainties in the mosaic CPR values,
because some of these variations on small scales are presumably the
result of varying surface properties. Thus, it can be
safely concluded that observational uncertainties in conjunction with
slopes and the bias introduced by parallax are not sufficient to
explain the measurements. This implies that there must be some
additional process responsible for changing the CPR in a systematic way
and that the interior surfaces of these polar anomalous craters are typically
different from their exteriors in more complicated ways than merely
having steeper slopes.

\subsection{The radial variation of CPR}\label{ssec:cprr}

Having determined that the angle of incidence is not solely
responsible for the differences between anomalous crater interiors and
exteriors, the challenge shifts to trying to determine what other
factors are affecting the CPR. Figure~\ref{fig:anglea} shows how the
median pixel
CPR varies with angle of incidence for different radial ranges both
inside and outside the anomalous craters. The pixels are placed into
the different radial bins based on their rectified positions within
the crater. For all different radial
ranges the shape of the median CPR variation with angle of incidence
is similar. Only the amplitude changes with radius.
The central region of the typical
crater has CPR values that are indistinguishable from those of pixels
in the crater exterior with $1.2<r/r_{\rm c}<1.5$. Out to
$r/r_{\rm c} \sim 0.8$, the CPR at a given angle of incidence increases
systematically with increasing radius. Inaccuracies in determining the
precise crater locations may scramble any trends at radii around
$r_{\rm c}$, but there is a sharper drop in the CPR outside the
crater edge than is seen inside the crater. 
No difference is seen in the results shown in Figure~\ref{fig:anglea}
when the anomalous crater sample is split in half either by crater
radius or latitude. The increased CPR at any
given angle of incidence seems to increase with increasing local
slope. At radii satisfying $0.5\lsim r/r_{\rm c}\lsim 1$,
where the CPR is largest for a given angle of incidence, the
azimuthally-averaged slopes are typically $\sim 25^\circ$. However,
the inaccuracy in the alignment 
of CPR and DEM maps and the relatively poor spatial resolution
preclude a more detailed comparison of CPR with local slope at present.

The corresponding results for the $9$ fresh craters are shown in
Figure~\ref{fig:anglef}. Wider bins in radius are used to prevent the
results becoming too noisy given the relatively small number of
fresh craters. The variation of CPR with angle of incidence is much
weaker than for the anomalous craters. Also, the radial variation,
while qualitatively similar to that seen for the anomalous craters, is
less pronounced. This is consistent with what one might expect from
a surface containing a uniform scattering of blocky ejecta behaving like
corner reflectors.

\begin{figure*}
\begin{center}
\includegraphics[trim=0.5cm 0cm 0cm 0.5cm,clip=true,width=2.1\columnwidth]{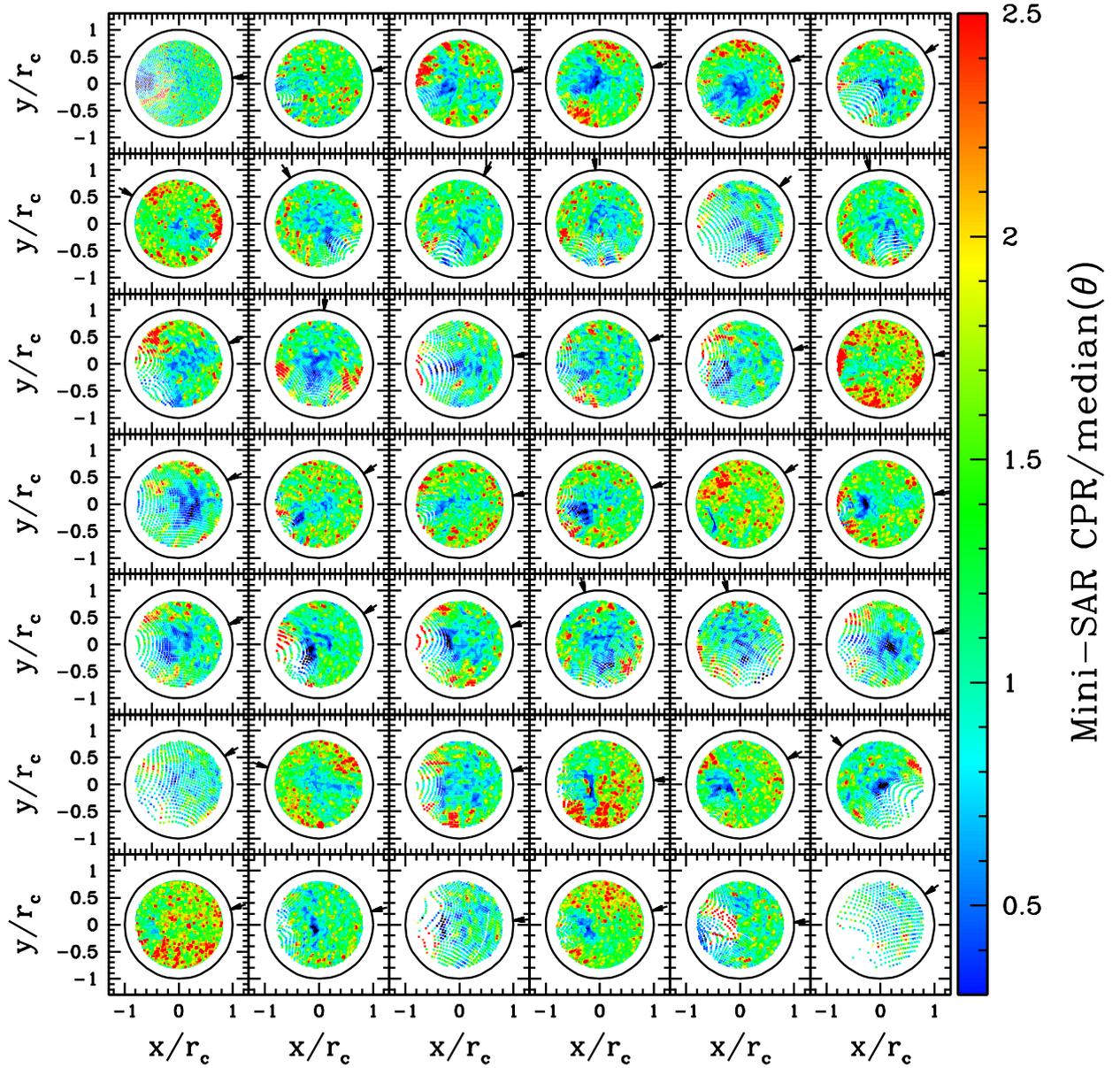} % 2-column
\end{center}
\vspace{-5cm}
\caption{Maps of Mini-SAR CPR/median CPR at that incidence angle for each of
the 42 craters. The craters are ordered as in
Figure~\ref{fig:allcdist} and the pixels are plotted at their
rectified locations, with north to the top. Median CPR as a function
of incidence angle is calculated for each of the craters
individually, using only the pixels with rectified radii having
$r/r_{\rm c}<0.8$. Black arrows show the azimuthal look direction inferred 
from the $S_1$ mosaic for each crater.}
\label{fig:manycor}
\end{figure*}

Maps of the variation of CPR relative to the typical value at each incidence
angle in each crater are shown in Figure~\ref{fig:manycor}. Although
the maps are quite heterogeneous, the relatively low CPR values tend to
be either in the crater centres or on the far wall as viewed by the
detector. Arrows show the direction in which each crater is viewed, as
determined from the high spots in the individual crater $S_1$
maps. Relatively high CPR values tend to be concentrated onto the
crater walls.
The median CPR values as a function of incidence angle are
determined from rectified pixels satisfying $r/r_{\rm c}<0.8$. This is done
to prevent errors arising from misalignments between the Mini-SAR
mosaic and the LOLA DEM. Near to the crater rim, the slopes change
rapidly, such that any misalignments between data sets would lead to
pixels being assigned very wrong incidence angles, biasing the
inferred CPR as a function of incidence angle. This effect may be
behind the slightly non-monotonic behaviour noted in
Figure~\ref{fig:anglea} for the radial bins adjacent to the rim.

\begin{figure}
\begin{center}
\includegraphics[trim=0cm 1cm 1cm 4cm,clip=true,width=0.95\columnwidth]{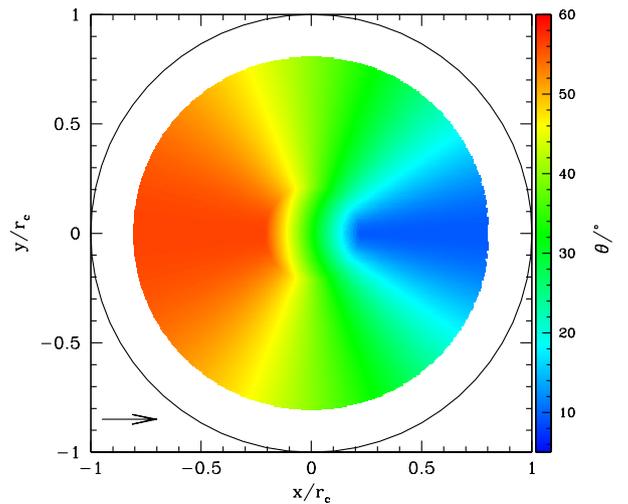}
\end{center}
\vspace{-2cm}
\caption{The distribution of incidence angle for the model crater
  considered in Section~\ref{ssec:slopes}. This shows which parts of a
  typical crater are viewed at the same angle of incidence, and
  represents a remapped version of Fig.~\ref{fig:modelr}. The detector
  is looking along the $+x$ direction at the model crater, as shown by
  the black arrow.}
\label{fig:look}
\end{figure}

Figure~\ref{fig:look} is included to help the interpretation of the
relative CPR maps in Figure~\ref{fig:manycor}. It shows how the angle
of incidence varies with position within the model crater used in
Section~\ref{ssec:slopes}, and is effectively just a rescaled version
of Figure~\ref{fig:modelr}. The comparison of 
local CPR with that at comparable angles of incidence, given in
Figure~\ref{fig:manycor} within each crater, is 
showing along a line of constant colour in Figure~\ref{fig:look}, with
the orientation set by the azimuthal look direction, 
where are the higher and lower values of CPR.

\section{Implications for the detection of water ice}\label{sec:disc}

\begin{figure}
\begin{center}
\includegraphics[trim=0cm 1cm 1cm 4cm,clip=true,width=0.95\columnwidth]{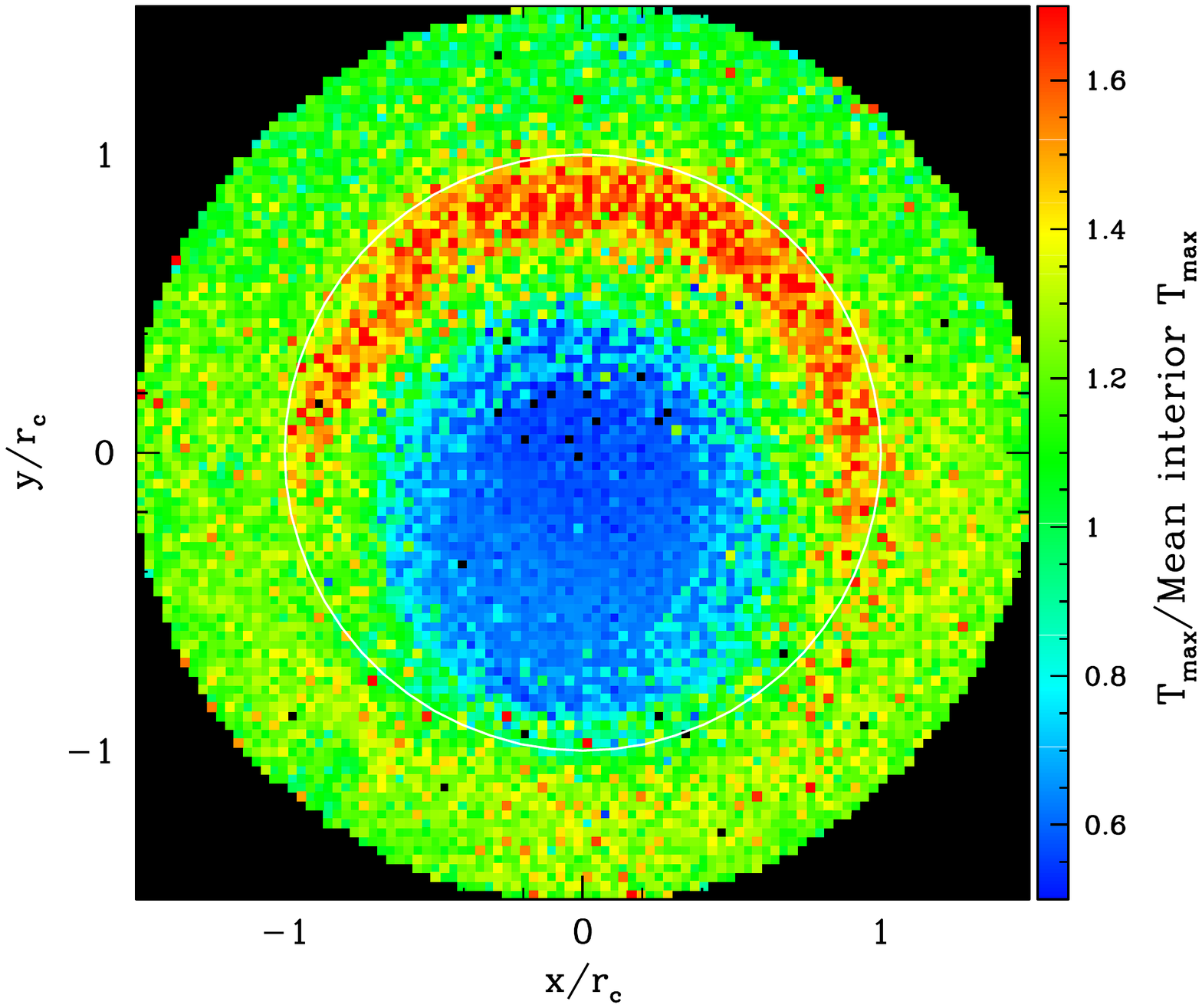}
\end{center}
\vspace{-2cm}
\caption{The stacked Diviner-inferred pixel $T_{\rm max}$ relative to
  the mean within each crater for the 33 anomalous craters. North is
  upwards, so the relatively cold part of the average crater is pole-facing.}
\label{fig:tempmap}
\end{figure}

The results in the previous section showed that high CPR regions
within polar anomalous craters, once angle of 
incidence effects are removed to the extent that is possible with the
data sets being used here, tend to be found on the steep crater
walls. This finding matches that of \cite{thom12} from their detailed study 
of Shackleton crater.
Figure~\ref{fig:tempmap} shows the stacked map
of the maximum temperature, $T_{\rm max}$, relative to the mean
maximum temperature within each crater, inferred from Diviner
measurements for the $33$ anomalous craters. 
For all craters, the largest interior $T_{\rm max}$ values exceed $290$K
and are found on the equator-facing walls, where direct sunlight can
occasionally be seen.
The stacked pole-facing slope and crater floor have the lowest
maximum temperatures, typically $70$K but ranging from $30-130$K, because
they only ever receive reflected sunlight. Given that surficial water
ice should be stable against sublimation for temperatures beneath
$\sim 100$K, one might well expect any water ice to be located in
these relatively cold regions within the craters.
This pattern of maximum temperatures is similar to that seen
in the average temperatures, and neither of them reflect the variation
of CPR, as might be expected if significant deposits of water ice
were responsible for the elevated interior CPRs in the anomalous polar
craters. 

It is possible that water ice could be insulated by a layer
of mantling regolith, in which case the CPR variations within
anomalous craters might not be expected to reflect those in the temperature.
Perhaps the central regions of craters are covered by too much
regolith for the radar to see underlying water ice. In contrast, the
steep crater sides should not be covered by deep regolith. However, in
these regions, the CPR variations still do not reflect the variations
in temperature determined using Diviner data.

\begin{figure}
\begin{center}
\includegraphics[trim=0cm 1cm 1cm 3cm,clip=true,width=0.95\columnwidth]{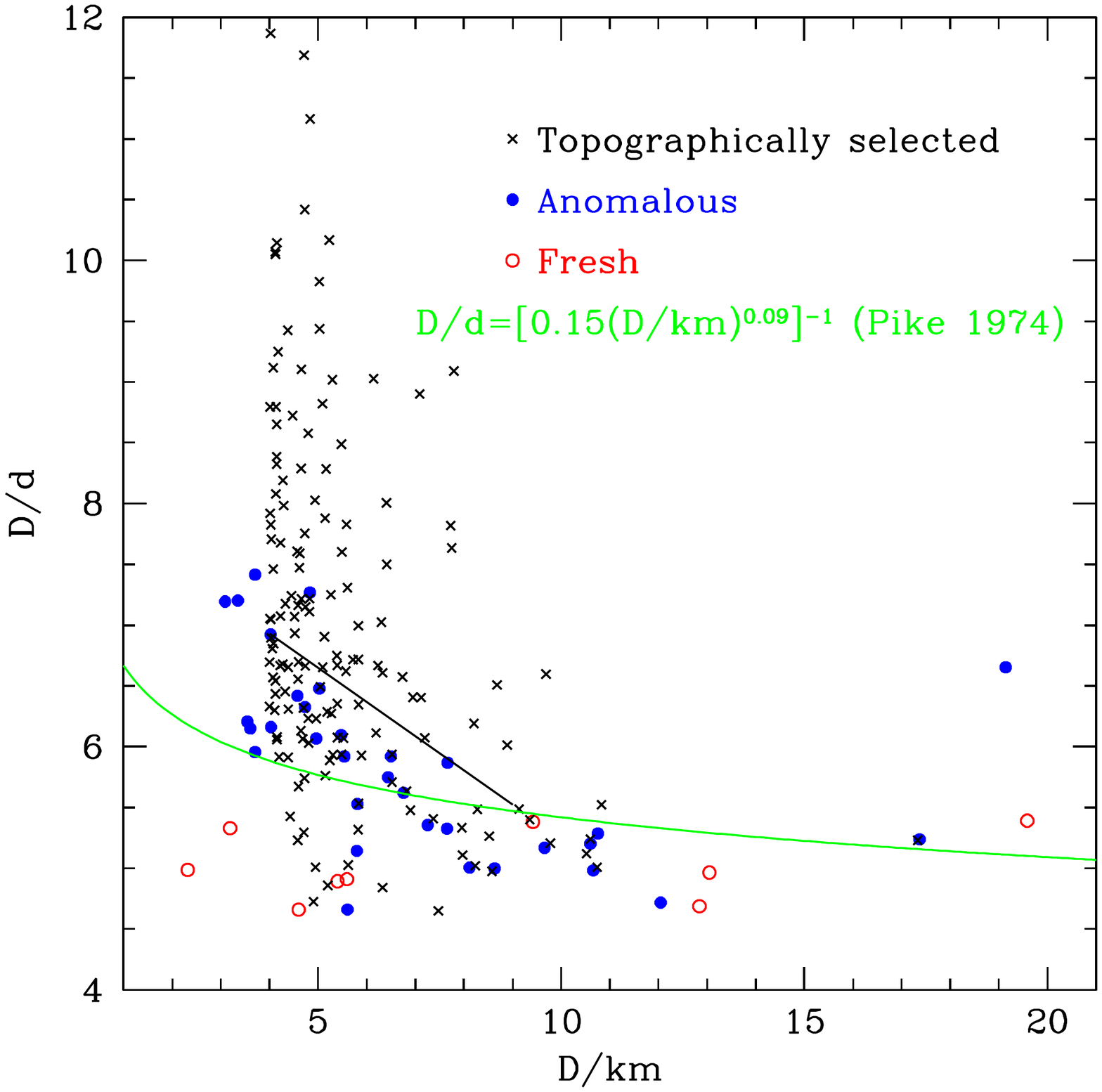}
\end{center}
\vspace{-2cm}
\caption{The variation of diameter-to-depth ratio ($D/d$) with crater
  diameter for the 33 anomalous craters (blue filled circles), 9 fresh
  craters (red open circles) and 154 topographically selected,
  isolated polar craters (black crosses). The black line represents
  the variation with diameter of the median of the black points, and
  the green line traces the relation given by Pike (1974) for fresh lunar craters.}
\label{fig:doverd}
\end{figure}

Using the set of $154$ topographically selected polar craters
described in the Appendix, one can
look at the diameter-to-depth ratios of the fresh and anomalous
craters relative to a set that have been found without reference to
their CPR properties. The mean diameter-to-depth ratios of the fresh
and anomalous craters are $D/d\sim 5.0$ and $5.9$
respectively. Increasing $D/d$ would be expected as craters age,
because the depths decrease over time while the diameters change little.
These measurements are therefore consistent with the picture of the anomalous
craters being older than the fresh ones. However, the topographically
selected craters have even larger $D/d$ values, with a mean of $\sim 7.0$.
Could these differences be driven by the crater diameter-to-depth
ratio varying with crater size? 
Figure~\ref{fig:doverd} shows the different crater sets as a function
of crater diameter. The solid black line represents the median $D/d$ for the
topographically selected craters binned into three different diameter
ranges, whereas the green line shows the relation found by
\cite{pike74} for a set of fresh lunar craters. It is clear that the
anomalous craters typically have lower 
diameter-to-depth ratios than the set of polar craters selected only on
topography. Under the assumption that $D/d$ is a proxy for crater age,
one therefore infers that the anomalous craters, while older than the
fresh ones, are still less mature than typical craters in the north polar
region. This is again suggestive that the effects of micrometeorite
bombardment
on the steep crater walls have not yet acted to remove all of the
rocks or roughness that give rise to high CPR values.

If micrometeoritic bombardment is isotropic and the blocky debris from
the crater forming impacts is weathered away at similar rates inside
and outside polar craters, then these results imply that processes are
preferentially acting on the steep slopes to refresh the near-surface
roughness to which the CPR is sensitive. This picture is consistent
with the findings of \cite{band11}, who use the thermal inertia
determined from Diviner measurements to infer rock abundances and
regolith thicknesses. They find extra rockiness on steep crater walls
relative to crater floors and crater exteriors, which is in
qualitative agreement with what is inferred in this study. Similarly,
\cite{fa13} use LROC images to show higher rock abundance interior to
craters relative to their exteriors. Furthermore, they find this extra
rockiness correlates with the difference between interior and exterior
CPR values, as measured by Mini-RF. Both the Diviner and LROC
rock abundances refer to objects that are at least $1-2$ m in size,
which is $\sim 10$ times the S-band radar wavelength. While there is
no guarantee that rockiness on these relatively large scales implies
roughness on scales more comparable with the radar wavelength,
the modelling of \cite{fa13} suggests that the larger rocks can
nevertheless provide a significant CPR enhancement through dihedral
reflections. 

If the anomalous craters do have high CPR as a result of differential
weathering of roughness, then the finding reported by \cite{spu13},
that the number density of anomalous craters at the poles greatly
exceeds that at lower latitudes, remains to be explained. This
apparent dependence on temperature is difficult to reconcile with the
indifference to local temperature of the CPR distribution within
anomalous polar craters. 
One would really like to start from the topographically-selected
crater sample and study the variation of CPR with crater morphology,
rather than starting from craters that have a particular CPR
distribution, as was done here and in previous work. Looking only at
CPR-selected craters can lead to  
a misleading impression of the population of craters as a whole. 
An orthorectified CPR mosaic, already
tied to the LOLA DEM, would be necessary to avoid
topographically-selected craters being ejected from the sample
if their CPR was insufficiently distinct for them to be detected via
their CPR, which has occurred in this study, as described in
Section~\ref{ssec:crat}. 

\section{Conclusions}\label{sec:conc}

The distribution of pixel CPR values inside and outside fresh craters
is largely independent of the angle of incidence with which  the lunar
surface is viewed. In contrast, for anomalous craters the angle of
incidence has a large impact on the CPR maps that result.
In these cases, counting pixels in SAR mosaics that have not been
rectified for the effect of parallax has the effect of biasing the
crater interior CPR pixel distribution to be dominated by observations
of the near wall, viewed at larger incidence angle. Consequently, the
mean interior crater CPR measured from an unrectified Mini-SAR map would
exceed that for the crater exterior even when the interior
and exterior surfaces have identical radar reflectivities (see
Figure~\ref{fig:modelcdist}).  

The typical variation of CPR with angle of incidence was measured
within the anomalous craters and used to make a model to quantify how using
unrectified mosaics will bias the distribution of pixel CPRs inside
the crater relative to that from just outside. While this
effect alone creates a sufficient change in the mean pixel CPR to
explain some of the anomalous craters, the additional scatter
required to recover the observed CPR distributions exceeds the
statistical uncertainties on the measurements. Therefore, the
CPR is also significantly affected by variations in the surface properties.

An additional variation with distance from the crater centre has also
been discovered, with the crater centre having CPR values like those
of the crater exterior, while larger CPR values at any given incidence
angle are found on the steeper parts of the crater walls. It is argued
that this variation of CPR with local slope, rather than local
temperature, suggests that it results from a variation
in the extent to which roughness is visible to the incident radar.
Steeper walls near the angle of repose may be less able to
sustain enough fine regolith to prevent the radar from seeing the
rougher rocks underneath or it could just be that ongoing weathering
produces more surface rocks or roughness on steeper slopes.

This argument is supported by the fact that anomalous craters,
while having larger diameter-to-depth ratios than fresh ones, are
typically steeper-sided than craters determined using a crater-finding
algorithm applied to the LOLA DEM. Assuming that the diameter-to-depth
represents a proxy for crater age, the anomalous craters are of
intermediate age. If surface roughness refreshed by mass-wasting on
steep slopes were responsible for the high CPR, then one would expect
anomalous craters to be of intermediate age, because fresh craters
have high CPR both inside and outside, whereas old craters do not
retain sufficiently steep sides for mass-wasting to continue to
promote sufficient surface roughness to cause high CPR. Thus, the
surface roughness explanation appears to pass this test.

Future analyses of the lunar SAR data should use properly controlled and
rectified CPR mosaics that are tied to the LOLA global DEM and take
into account explicitly the dependence of CPR on angle of
incidence.
The model of \cite{fa11}, while not including multiple scattering and
the CBOE, suggests that radar data will not be able to distinguish
between regolith with and without a few wt\% WEH, which is the level
that the LCROSS and LPNS results imply is the likely concentration. 
There is strong circumstantial evidence that the 
extractable information from the lunar SAR data will pertain to
surface or near-surface roughness rather than water ice. This should
provide fertile ground in conjunction with Diviner and LROC data sets
to learn about surface weathering as a function of local slope and
composition \citep{bell12}. 

\section*{Acknowledgments}
VRE thanks Randy Kirk for helpful comments. We would like to thank the
referees for their helpful comments.
This work was supported by the Science and Technology Facilities Council 
[grant number ST/F001166/1].
LT acknowledges the support of the LASER and PGG NASA
programs for funding this research.

\def\jgr{J. Geophys. Res. }
\def\grl{Geophys. Res. Lett. }
\def\nat{Nature }
\def\mnras{Mon. Not. R. Astron. Soc. }
\def\icarus{Icarus}
\def\pasp{Publ. Astron. Soc. Pac. }

\section*{References}

\bibliographystyle{elsarticle-harv}
\bibliography{mybib}

%\begin{thebibliography}
%
%\end{thebibliography}

\begin{appendix}

\section{Crater-finding algorithm}\label{app:meth}

The list of craters produced by \cite{head10} from the LOLA
topographical data consists of 5185 craters with radii of at least 
$10$ km distributed over the entire lunar surface. \cite{sala12}
supplemented this with additional craters found using a predominantly
automated detection
algorithm that was based on the LOLA DEM. Their crater catalogue contained
60645 objects and is the most complete to radii of $4$ km.
For the purpose of this study, even smaller craters in the vicinity
of the lunar north pole are of interest, and the desire is to produce
craters with representative diameter-to-depth ratios. Thus, an
algorithm has been developed to find simple craters 
with radii in the range $2\leq r_{\rm c}/{\rm km}\leq 10$ using the
LOLA north polar stereographic digital elevation map.

The crater-finding algorithm consists of two main stages. First, by
placing `water' on the surface and letting it drain downhill to create
puddles, a set of potential crater centres are found. The
amount of water in each puddle reflects the area from which it came
and hence provides an estimate of the radius of
the potential crater. Secondly, in the vicinity of each potential
crater, the Laplacian of the topography is filtered to search for
circularly symmetric patterns with a concave centre surrounded by a
convex rim. The details of these two parts of the algorithm are
described in the following subsections.

\subsection{Finding crater candidates}\label{app:water}

Candidates for crater centres are found using a hydrological algorithm
that is a simplified version of those described by \cite{hydro1} and
\cite{hydro2}. 
A smoothed version of the LOLA polar stereographic $80$ m DEM is used.
The smoothing suppresses small scale depressions that might
otherwise prevent `water' from draining further into larger
depressions. It also removes candidate tiny craters that might be within other
craters, which would consequently fail the isolation criterion described
in the next section and be jettisoned from the sample.
A single smoothing entails replacing each altitude with a value that is $1/4$
of the original value plus $1/8$ of each of the values in the $4$
adjacent pixels, plus $1/16$ times the values in the diagonally
adjacent pixels. Given that craters in the radius range $2-10$ km are
being considered here, 3 smoothings of the DEM are used.

An amount of `water' proportional to the pixel area is
placed into each pixel in the smoothed digital elevation map and this is allowed
to run downhill using the following iterative method.
Each pixel with none of its 8 neighbours being higher and containing
water, distributes its water to neighbouring pixels that are lower than
it. The water is distributed to the $N$ lower neighbouring
pixels in proportion to the gradient in their direction. Thus, the
fraction of water sent to the $i$th lower neighbour is given by
\begin{equation}
f_i=\frac{|\underline{\nabla}_i|}{\sum_{j=1}^{N}|\underline{\nabla}_j|},
\end{equation}
where $\underline{\nabla}_i$ represents the gradient in the direction of
the $i$th neighbour. This draining is repeated until no pixels with
lower neighbours contain any water, at which point the set of `wet' pixels
defines the centres of crater candidates, with the amount of
water providing an estimate of the potential crater radius under the
assumption that it came from a circular patch of the surface.

\subsection{Confirming craters}\label{app:filter}

For the purpose of this study, there is no need to have a complete
sample of craters, merely one that is representative of the
diameter-to-depth ratios of craters as a whole. Thus, for simplicity,
only isolated crater candidates are retained for further
consideration. Isolation is defined as having no other crater
candidate within one candidate crater radius from the candidate crater
centre. This yields a set of $\sim 68000$ candidate isolated craters of
all radii at ${\rm latitude}>80^\circ$.
These candidates are then filtered to refine the centres and radii and
determine a statistic related to how much they match a simple crater
in their topographic profile.

The Laplacian of the DEM in the vicinity of each of these potential
craters is filtered using a compensated filter of the form
\begin{equation}
w(r)=\left \{
\begin{array}{ll}
N_{\rm ring}/N_{\rm cen} & {\rm if~~~} r<0.6r_{\rm c,test},\\
-1 &{\rm if~~} |r-r_{\rm c,test}|\leq 40 {\rm m,}\\
0 &{\rm otherwise,}
\end{array}
\right.
\end{equation}
where $r_{\rm c,test}$ is the crater radius being tested, $N_{\rm cen}$
is the number of $80$ m pixel centres lying within a disc of radius
$0.6r_{\rm c,test}$ and $N_{\rm ring}$ is the number of pixel centres
within an annulus one pixel wide having mean radius equal to $r_{\rm c,test}$.
Crater radii are tested in the range $0.5-1.5$ times the value inferred
from the amount of water gathered by each candidate. This filter picks
out regions that have a concave disc of surface surrounded by a convex
rim-like structure. The
pixels within which the maximum filtered Laplacian values are found
for each tested crater radius provide the most likely crater centres
for those test radii.

To determine which tested radius produces the best overall match, a
significance of the value of the filtered Laplacian is
defined. Applying the filter to a random part of the Laplacian map
inferred from the DEM would give rise to a distribution of filter
values. This can be treated as a random walk with a step size of the
rms Laplacian weighted by the rms step size of the filter. Using this
to normalise the filtered Laplacian values around the candidate crater
centre gives a significance for each candidate crater. This value is used to
determine the best test radius. Each candidate with a significance, $S$, (of
the filtered Laplacian relative to that expected from a random walk)
of at least $S_{\rm min}=15$ is deemed to be a detected crater.

\subsection{The set of polar craters}\label{app:crat}

\begin{figure}
\begin{center}
\includegraphics[trim=0cm 1cm 1cm 3cm,clip=true,width=0.95\columnwidth]{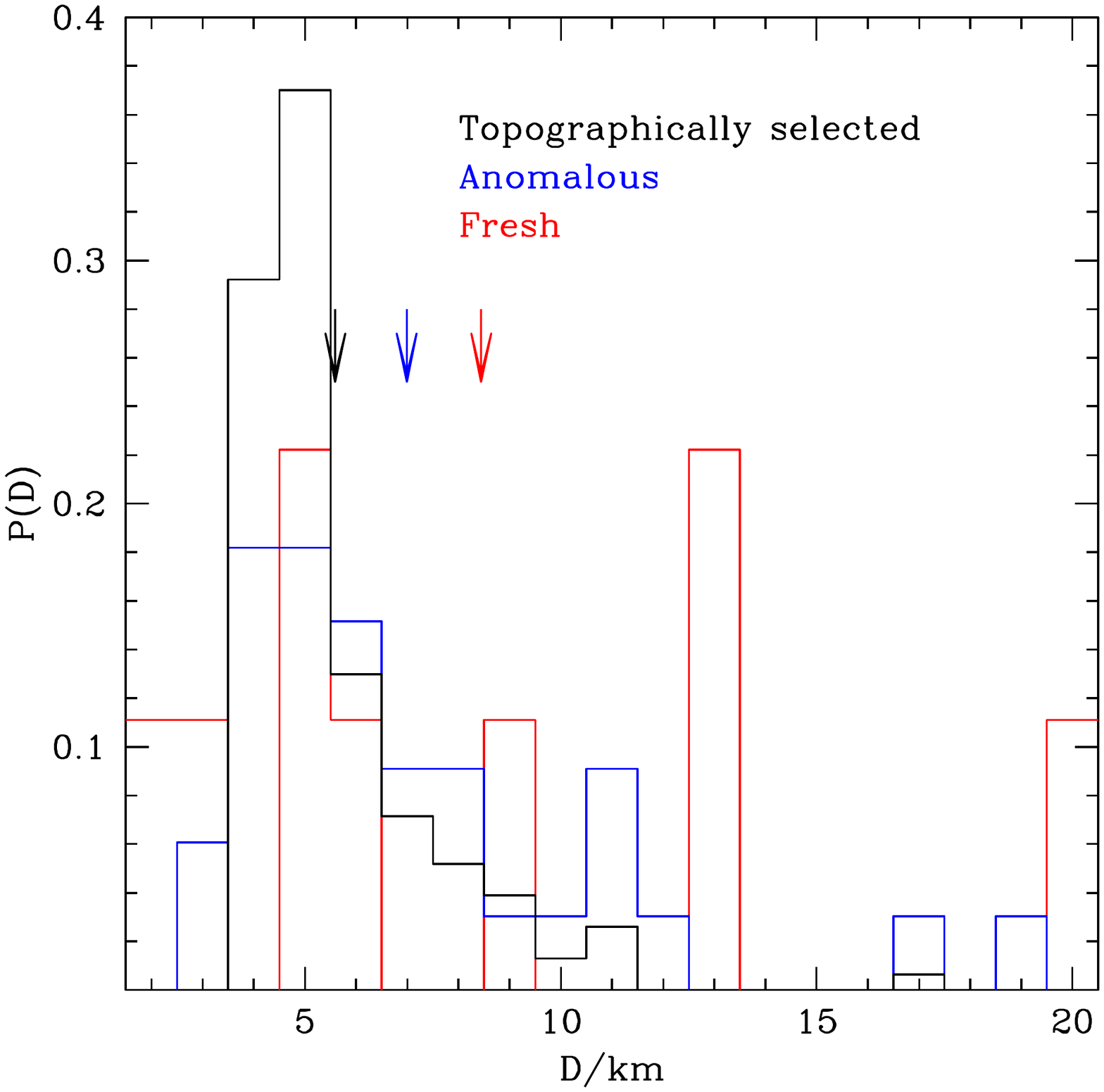}
\end{center}
\vspace{-2cm}
\caption{The probability distributions of the crater diameters for the
  three different sets of craters: $154$ topographically selected (black),
  $9$ fresh (red) and $33$ anomalous (blue). Coloured arrows show the mean
  diameters in each sample.}
\label{fig:app1}
\end{figure}

The algorithm described above yields $154$ craters with
latitude greater than $80^\circ$.
Table~A1 contains a list of the centres and radii of
these north polar, isolated craters, and Figure~\ref{fig:app1} shows their
distribution with diameter. Figure~\ref{fig:doverd} plots the
dependence of the crater diameter-to-depth ratios on diameter,
illustrating how these topographically selected craters typically have
shallower profiles than either the fresh or anomalous craters studied
by \cite{spu10}.

\begin{figure}[H]
\begin{center}
\includegraphics[trim=0cm 1cm 1cm 3cm,clip=true,width=0.95\columnwidth]{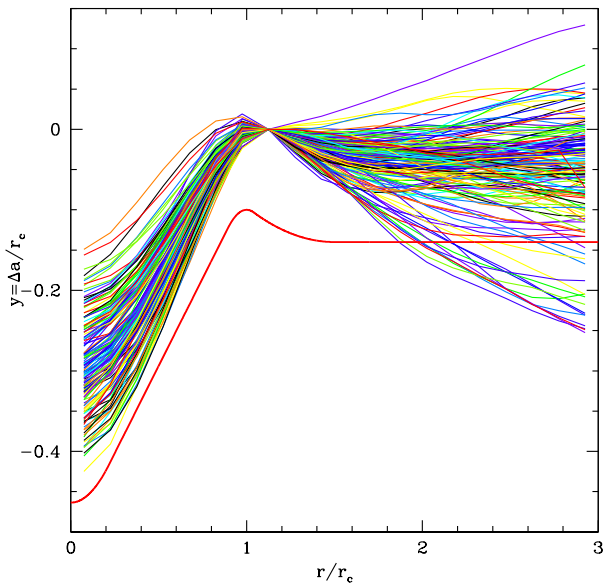}
\end{center}
\vspace{-2cm}
\caption{The azimuthally averaged height profiles, scaled by crater
  radius, for the $154$ 
  topographically selected craters. Each radius is rescaled by the
  crater radius, $r_{\rm c}$, whereas the scaled height is plotted relative
  to the value at $r/r_{\rm c}=1.1$. The bold red line shows the
  profile for the model crater used in Section~\ref{ssec:slopes},
  offset vertically by $0.1$ for clarity.}
\label{fig:app2}
\end{figure}

The choice of $S_{\rm min}$ feeds into the inferred
diameter-to-depth ratio of the resulting crater catalogue, because
deeper craters better match the filter shape than shallower
ones. Thus, increasing $S_{\rm min}$ from $15$ to $20$ decreases the
number of craters from $154$ to $108$, and the diameter-to-depth
ratio from $7.0$ to $6.3$. However, the lower threshold of 
$S_{\rm min}=15$ still produces a set of azimuthally symmetric
depressions with convex rims that are crater-like. 
Figure~\ref{fig:app2} shows the azimuthally-averaged height profiles,
scaled by crater radius,
of all $154$ craters with $S>15$. The diversity of depths reflects
the range of diameter-to-depth values for the selected craters, and
it is apparent that each of the
craters possesses both a central depression and a convex rim.
\begin{table*}\label{tab:app}
\begin{center}
\caption{Radii and locations for the $154$ topographically selected isolated
craters. Longitudes and latitudes are given in degrees.}
%\footnotesize{
\scriptsize{
\begin{tabular}{l|c|c|c|c|c|c|c|c|c|c|c}\hline
Crater \# & $r_{\rm c}$/km & (lat,lon) & Crater \# & $r_{\rm c}$/km & (lat,lon) &
Crater \# & $r_{\rm c}$/km & (lat,lon) & Crater \# & $r_{\rm c}$/km & (lat,lon) \\ \hline
1 & 2.4 & 80.01, -21.4 & 2 & 2.7 & 80.01, 31.8 & 3 & 2.1 & 81.12, -21.4  & 4 & 2.9 & 81.35, 19.0 \\
5 & 4.9 & 81.65, -23.9 & 6 & 2.6 & 82.26, 11.7 & 7 & 2.4 & 81.84, 28.2  & 8 & 2.3 & 81.49, -32.9 \\
9 & 2.7 & 81.85, 29.2 & 10 & 2.1 & 81.83, -31.1 & 11 & 2.1 & 80.07, -46.6  & 12 & 2.2 & 81.98, -34.3 \\
13 & 3.9 & 82.65, 26.7 & 14 & 2.0 & 83.29, -13.7 & 15 & 2.7 & 82.07, -37.1  & 16 & 2.5 & 83.06, 24.5 \\
17 & 8.7 & 80.26, -50.1 & 18 & 2.0 & 82.49, -34.2 & 19 & 2.8 & 83.87, -7.4  & 20 & 5.3 & 83.76, -13.9 \\
21 & 2.5 & 82.68, -37.4 & 22 & 2.4 & 84.12, 15.7 & 23 & 3.6 & 80.80, 53.7  & 24 & 2.0 & 84.14, -21.7 \\
25 & 2.2 & 84.18, -20.4 & 26 & 2.3 & 84.64, -6.2 & 27 & 2.3 & 80.01, 61.6  & 28 & 2.0 & 81.87, -56.8 \\
29 & 2.1 & 81.43, 59.7 & 30 & 4.3 & 80.16, -66.1 & 31 & 3.7 & 80.32, 65.9  & 32 & 3.7 & 85.78, 25.2 \\
33 & 2.9 & 85.91, -27.7 & 34 & 2.1 & 83.64, -55.7 & 35 & 3.5 & 80.46, -68.7  & 36 & 2.8 & 81.19, -68.2 \\
37 & 4.8 & 83.89, -57.4 & 38 & 2.3 & 84.88, -50.7 & 39 & 2.6 & 83.94, -59.3  & 40 & 2.8 & 85.75, 43.6 \\
41 & 2.2 & 81.40, 69.7 & 42 & 2.1 & 84.57, -56.7 & 43 & 2.4 & 85.30, -52.3  & 44 & 2.3 & 81.08, 71.8 \\
45 & 5.4 & 86.99, 28.6 & 46 & 2.3 & 85.79, 54.0 & 47 & 3.5 & 81.85, 72.8  & 48 & 3.2 & 87.12, -33.4 \\
49 & 2.0 & 86.89, -45.6 & 50 & 2.0 & 87.52, -29.3 & 51 & 2.5 & 87.69, 30.8  & 52 & 2.6 & 83.91, 72.4 \\
53 & 2.1 & 84.51, -70.6 & 54 & 2.7 & 81.47, -77.8 & 55 & 2.4 & 82.80, 75.5  & 56 & 2.9 & 87.97, 29.9 \\
57 & 2.7 & 86.64, 58.5 & 58 & 2.3 & 88.08, -27.8 & 59 & 2.6 & 84.90, -71.5  & 60 & 2.1 & 88.22, -26.0 \\
61 & 3.4 & 88.26, 25.2 & 62 & 3.4 & 81.50, -79.8 & 63 & 2.7 & 88.08, 39.9  & 64 & 3.1 & 87.92, 57.1 \\
65 & 3.1 & 87.66, 63.2 & 66 & 2.4 & 85.59, 76.9 & 67 & 4.7 & 87.36, 68.0  & 68 & 2.2 & 86.01, 76.0 \\
69 & 2.5 & 86.65, 73.7 & 70 & 3.2 & 85.75, 78.1 & 71 & 2.1 & 87.81, -66.8  & 72 & 2.5 & 88.75, 47.0 \\
73 & 2.3 & 82.66, -83.6 & 74 & 3.2 & 81.56, -84.6 & 75 & 3.3 & 88.19, 63.4  & 76 & 2.6 & 85.56, 79.5 \\
77 & 2.6 & 88.96, -45.1 & 78 & 3.9 & 88.05, 68.4 & 79 & 2.8 & 82.71, -87.1  & 80 & 2.9 & 81.22, 88.4 \\
81 & 2.4 & 83.32, 88.2 & 82 & 4.1 & 87.13, -86.3 & 83 & 2.6 & 85.97, 88.1  & 84 & 2.0 & 89.64, -108.8 \\
85 & 2.6 & 86.27, 94.0 & 86 & 2.0 & 86.89, 96.4 & 87 & 2.3 & 83.65, 93.7  & 88 & 2.0 & 88.17, 112.0 \\
89 & 2.0 & 87.69, 107.9 & 90 & 2.4 & 87.83, 113.0 & 91 & 2.5 & 85.43, 101.1  & 92 & 3.4 & 87.41, 110.0 \\
93 & 2.7 & 80.43, -99.5 & 94 & 2.0 & 88.41, -177.9 & 95 & 2.1 & 84.87, -109.0  & 96 & 2.1 & 81.87, -102.2 \\
97 & 2.1 & 82.54, 105.1 & 98 & 2.4 & 83.09, -106.4 & 99 & 2.1 & 87.24, 135.7  & 100 & 2.5 & 83.74, -108.7 \\
101 & 3.6 & 83.35, -108.3 & 102 & 2.5 & 84.00, -110.7 & 103 & 2.1 & 81.40, -104.7  & 104 & 2.9 & 84.55, -114.5 \\
105 & 3.9 & 82.41, 107.5 & 106 & 2.0 & 87.66, 172.3 & 107 & 2.5 & 84.72, -116.9  & 108 & 2.2 & 81.45, 106.2 \\
109 & 2.4 & 85.94, -129.2 & 110 & 3.1 & 80.51, 107.7 & 111 & 2.1 & 81.85, 110.8  & 112 & 2.4 & 86.64, 152.8 \\
113 & 2.2 & 85.63, -134.4 & 114 & 2.4 & 84.62, 125.5 & 115 & 5.4 & 82.63, 116.0  & 116 & 2.0 & 85.15, 132.9 \\
117 & 2.3 & 85.24, 136.3 & 118 & 2.4 & 81.54, -114.6 & 119 & 2.6 & 81.01, 113.5  & 120 & 2.0 & 84.44, -130.7 \\
121 & 2.4 & 84.00, -127.4 & 122 & 3.2 & 84.49, -132.4 & 123 & 2.2 & 86.18, -177.6  & 124 & 2.2 & 85.85, -157.1 \\
125 & 4.4 & 81.08, 115.8 & 126 & 2.0 & 83.37, 126.1 & 127 & 4.1 & 80.93, -115.6  & 128 & 2.2 & 81.45, 118.2 \\
129 & 4.0 & 83.10, 129.9 & 130 & 3.2 & 83.73, -135.3 & 131 & 4.0 & 80.45, -122.7  & 132 & 4.1 & 82.24, -134.2 \\
133 & 4.3 & 84.05, -156.4 & 134 & 3.3 & 80.17, -124.6 & 135 & 2.3 & 80.58, 126.4  & 136 & 2.6 & 84.10, -163.4 \\
137 & 2.1 & 83.22, 147.8 & 138 & 2.9 & 83.99, 174.0 & 139 & 2.1 & 84.00, -176.2  & 140 & 2.3 & 82.90, 150.9 \\
141 & 4.6 & 82.46, 145.7 & 142 & 5.3 & 81.15, 137.7 & 143 & 2.3 & 80.31, -135.4  & 144 & 2.3 & 80.74, 146.0 \\
145 & 2.3 & 80.97, -161.1 & 146 & 2.1 & 81.29, 171.3 & 147 & 2.1 & 80.97, -162.7  & 148 & 4.3 & 80.46, -155.1 \\
149 & 2.7 & 80.38, -156.1 & 150 & 2.6 & 80.71, 164.1 & 151 & 2.1 & 80.96, 172.6  & 152 & 2.9 & 80.27, 158.6 \\
153 & 2.8 & 80.84, 173.3 & 154 & 2.1 & 80.18, 176.4  \\
\hline
\end{tabular}}
\end{center}
\end{table*}

\end{appendix}

%\end{article}

\end{document}